\documentclass[journal]{IEEEtran}

\usepackage{graphicx}

\usepackage{dblfloatfix}
\usepackage{rotating}
\usepackage{booktabs}
\usepackage{multirow}
\usepackage[table]{xcolor}
\usepackage{longtable}
\usepackage{lscape}
\usepackage{amsmath, amsthm, amssymb}
\usepackage{url}
\usepackage{lipsum}
\usepackage[T1]{fontenc} 
\usepackage{enumitem}
\usepackage[utf8]{inputenc}

\usepackage{url}

\usepackage[caption=false,font=footnotesize]{subfig}

\usepackage{xcolor}

\begin{document}

\title{A SDN/OpenFlow Framework for Dynamic Resource Allocation based on Bandwidth Allocation Model}

\author{Eliseu~Torres,
        Rafael~Reale,
        Leobino~Sampaio,
        and~Joberto~Martins,~\IEEEmembership{Senior~Member,~IEEE}
\thanks{MSc. Torres, E. is with Universidade Federal da Bahia - UFBA, Salvador, Brazil. (e-mail: eliseu.torres@ufba.br)}
\thanks{Prof. Dr. Reale, R. is with Instituto Federal de Educação - IFBA, Valença, Brazil. \protect  (e-mail: reale@ifba.edu.br)}
\thanks{Prof. Dr. Sampaio, L. is with Universidade Federal da Bahia - UFBA, Salvador, Brazil. \protect (e-mail: leobino@ufba.br)}
\thanks{Prof. Dr. Martins, J. is with Salvador University - UNIFACS, Brazil. \protect (e-mail: joberto.martins@gmail.com)}
}

\markboth{IEEE Latin America Transactions,~Vol.~18, No.~5, May~2020}%
{Shell \MakeLowercase{\textit{et al.}}: Bare Demo of IEEEtran.cls for IEEE Journals}

\maketitle

\begin{abstract}
The communication network context in actual systems like 5G, cloud and IoT (Internet of Things), presents an ever-increasing number of users, applications and services that are highly distributed with distinct and heterogeneous communications requirements. Resource allocation in this context requires dynamic, efficient and customized solutions and Bandwidth Allocation Models (BAMs) are an alternative to support this new trend. This paper proposes the BAMSDN (Bandwidth Allocation Model through Software-Defined Networking) framework that dynamically allocates resources (bandwidth) for a MPLS (MultiProtocol Label Switching) network using a SDN (Software-Defined Networking)/OpenFlow strategy with BAM. The framework adopts an innovative implementation approach for BAM systems by controlling the MPLS network using SDN with OpenFlow. Experimental results suggest that using SDN/OpenFlow with BAM for bandwidth allocation does have effective advantages for MPLS networks requiring flexible resource sharing among applications and facilitates the migration path to a SDN/OpenFlow network.
\end{abstract}


\IEEEpeerreviewmaketitle

\section{Introdução}
\label{sec:introduction}

\IEEEPARstart{C}{enários} de redes de comunicação atuais e emergentes, tais como 5G, nuvem, redes ópticas, redes MPLS (\textit{MultiProtocol Label Switching}) e internet das coisas (IoT -- \textit{Internet of Things}), são marcados pela grande variedade e distribuição de usuários, aplicações e serviços, com requisitos de comunicação e de qualidade (SLA -- \textit{Service Level Agreement}, QoS -- \textit{Quality of Service}, QoE -- \textit{Quality of Experience}) heterogêneos. Tais cenários apresentam rápido crescimento, como consequência natural de novos requisitos dos usuários, o que tem levado à gerência de redes e sistemas objetivos bastante desafiadores. Mais especificamente, busca-se uma rede capaz de se adequar dinamicamente e eficientemente às variações de contexto existentes, o que torna a alocação dinâmica, flexível e customizada de recursos essencial \cite{rachkidi_cloud_2017,zuo_fast_2015}. 

A alocação dinâmica é necessária, por exemplo, para lidar com a variação do tráfego de entrada das redes \cite{fan_dynamic_2016}. Já a flexibilidade é mais ligada ao fato que a gerência deve, sempre que possível, otimizar o uso da rede. Por fim, a customização está diretamente relacionada com a perspectiva de utilização de soluções de alocação de recursos que possam ser utilizadas com algum tipo de configuração em redes heterogêneas, p.ex., IoT e nuvem.

Os modelos de alocação de banda (BAM - \textit{Bandwidth Allocation Models}) possuem os atributos para resolver esses três requisitos (dinamismo, flexibilidade e customização), pois permitem a definição de classes de aplicações e o controle da distribuição dos recursos entre as classes~\cite{ martins_uma_2015, trivisonno_network_2015, pistek_-mar:_2015}. Por outro lado, no processo de alocação, implementações atuais de BAMs não exploram as questões de otimização tratadas pela academia e indústria nos últimos anos, que sugerem um grande número de algoritmos e heurísticas de engenharia de tráfego~\cite{onali_traffic_2008}. Portanto, o principal desafio atual está em viabilizar a alocação de recursos nos equipamentos, com a otimização sendo tratada pelo BAM, de forma que seja levada em conta a distribuição dos recursos e sua efetivação de forma dinâmica. Na prática, tal desafio torna-se ainda mais relevante quando se leva em consideração a heterogeneidade da rede, distribuição geográfica, características dos equipamentos e dinamicidade do tráfego.  

O paradigma das redes definidas por software (SDN - \textit{Software-Defined Networking})~\cite{kreutz_software-defined_2014} traz novas perspectivas. O plano de controle, logicamente centralizado, permite que a visão global da rede seja utilizada para explorar a definição de rotas e os equipamentos existentes nos enlaces de rede, ao invés de limitar-se ao escopo individual de cada equipamento. Ao implementar BAMs através de SDN, é possível superar dificuldades inerentes ao problema de alocação de recursos, sobretudo em ambientes caracterizados pela heterogeneidade e distribuição. Por tais motivos, o presente trabalho apresenta o arcabouço BAMSDN (\textit{Bandwidth Allocation Model through Software-Defined Networking}) para a exploração dinâmica e flexível de recursos de redes MPLS (\textit{MultiProtocol Label Switching}) através de uso conjunto de BAMs e SDN. A partir da programabilidade e visão global do plano de controle da rede, utilizadas em conjunto com o BAM, o arcabouço busca adequar a disponibilidade dos recursos de rede à demanda dos seus usuários.  A validação do arcabouço em relação à sua capacidade de realizar uma reconfiguração dinâmica e flexível de banda será feita experimentalmente.

Na parte seguinte deste artigo, a Seção~\ref{sec:firstpage} resume os aspectos fundamentais dos BAM e os trabalhos relacionados. A Seção~\ref{sec:bamsdn} descreve o arcabouço BAMSDN com seus componentes e operação. A Seção~\ref{sec:experimentacao} trata da experimentação que visou subsidiar a avaliação do arcabouço proposto. A Seção~\ref{sec:analise} faz uma análise dos resultados obtidos. Por fim, a Seção~\ref{sec:conclusao} apresenta as considerações finais e trabalho futuro.

\section{BAMs e Estratégia SDN/ OpenFlow: Aspectos Básicos para a Alocação de Banda} \label{sec:firstpage}

O arcabouço BAMSDN proposto e descrito na próxima seção reúne dois elementos que se superpõem de forma vantajosa para a alocação de banda com dinamismo e flexibilidade em redes MPLS: i) Um modelo de alocação de banda (BAM); e ii) Uma estratégia SDN com o OpenFlow como protocolo para o controle da operação da rede. Esta seção apresenta os princípios básicos dos BAM e discute os trabalhos relacionados.

\subsection{BAMs: Modelos de alocação de banda - Operação básica e configuração}

Modelos de Alocação de Banda (BAM) são uma estratégia de alocação de recurso, inicialmente focada no recurso largura de banda (\textit{bandwidth}), que permite três ações principais para a alocação do recurso banda: i) um planejamento e agrupamento dos seus usuários em classes de serviços com requisitos comuns (QoS, SLA ou outro critério/ requisito do usuário ou aplicação); ii) uma definição da quantidade de recursos necessária (alocável) por classe (BC - \textit{Bandwidth Constraint}); e iii) uma estratégia ou comportamento (\textit{behavior}) para a alocação dos recursos com compartilhamento ou não entre os recursos associados por classe  \cite{martins_uma_2015}. 

O planejamento do agrupamento de usuários e aplicações em classes de tráfego (CT) juntamente com a definição de quanto de recurso (banda) é alocada para cada classe é um aspecto de configuração dos BAM. Obviamente, esta configuração pode ser alterada de forma dinâmica durante a operação da rede e, neste sentido, suporta um dinamismo na alocação da banda disponível (alocação dinâmica de banda), por exemplo, em relação à demanda de tráfego, requisitos de operação (SLA, QoS, QoE) ou número de usuários ativos na rede \cite{martins_uma_2015}. Um segundo aspecto dinâmico dos BAM é o seu comportamento (\textit{behavior}) na alocação de banda entre as classes de serviço. Neste caso existem diferentes tipos de compartilhamento nos diferentes tipos de BAM (MAM, RDM, ATCS, GBAM) que também são utilizados visando a dinamicidade, customização e eficiência da alocação de recursos em relação às demandas dos usuários da rede. 

Existem três modelos básicos para os BAM: i) MAM (\textit{Maximum Allocation Model}) \cite{faucher_maximum_2005}, ii) RDM (\textit{Russian Dolls Model}) \cite{pinto_da_costa_neto_algoritmos_2008} e iii) AllocTC-Sharing (ATCS) \cite{reale_alloctc-sharing:_2011}. Além destes, existem vários BAM que são efetivamente híbridos ou propõem a estratégia dos BAM básicos com algumas modificações \cite{pistek_-mar:_2015} e um modelo generalizado de alocação da banda (GBAM - \textit{Generalized Bandwidth Allocation Model}) \cite{martins_g-bam:_2014} que permite a configuração de todos os comportamentos possíveis.

Em resumo, o modelo básico MAM aloca banda sem permitir o compartilhamento de recurso entre as classes de tráfego (CTs). O modelo RDM permite o compartilhamento de banda das classes de tráfego mais prioritárias para classes de tráfego menos prioritárias (estratégia HTL - \textit{High-to-Low}). Já o modelo ATCS permite de forma oportunista o compartilhamento generalizado de banda entre todas as classes de tráfego prioritárias e não prioritárias (estratégias HTL e LTH- \textit{Low-to-High}) \cite{martins_uma_2015}. 

O modelo de alocação de banda se aplica para todos os enlaces (\textit{links}) na trajetória (\textit{path}) do LSP (\textit{Label Switched Path})/MPLS (\textit{MultiProtocol Label Switching)} sendo alocado. Assim sendo, o BAM funciona como um tipo de procurador  (\textit{broker}) para a alocação de banda, autorizando a criação de um LSP (\textit{setup grant}), rejeitando o pedido de LSP (\textit{setup deny}) ou autorizando a criação de um LSP com a preempção e/ou devolução de outro(s) quando existe compartilhamento entre classes de tráfego.

\subsection{Explorando de forma dinâmica e flexível a alocação de banda com os BAM} \label{Problema}

No BAMSDN, o BAM tem dois papeis fundamentais em relação à alocação de banda: i) o BAM na sua operação normal distribui os recursos limitados existentes (banda) entre um conjunto de classes de aplicações configuradas; e ii) o BAM, através da reconfiguração dos limites de banda por classe de aplicação (BC), permite uma otimização na alocação do recurso em função do perfil de tráfego dinâmico da rede.

A reconfiguração de BCs é um elemento adicional importante para a flexibilidade na alocação dos recursos. Em resumo, se o perfil da rede muda, o arcabouço BAMSDN pode ajustar as alocações de banda por classe de aplicações de forma a melhor refletir tanto a dinâmica da demanda como as necessidades dos usuários. Em termos técnicos, a reconfiguração de BC é o processo no qual alteramos as configurações iniciais das restrições de banda (BC) definidas para uma ou mais CTs em tempo de execução.

Em \cite{reale_preliminary_2014} foram propostas duas abordagens para reconfiguração de BCs dos modelos de alocação de bandas (BAMs): i) a abordagem \textit{hard}; e ii) a abordagem \textit{soft}. A abordagem \textit{hard} tem como premissa a reconfiguração imediata dos BCs mesmo que esta reconfiguração implique na liberação abrupta de recursos sendo utilizados. Em termos operacionais, a abordagem \textit{hard} pode gerar, por exemplo, preempções numa rede MPLS. Já a abordagem \textit{soft} tem como premissa minimizar os possíveis impactos resultantes na reconfiguração dos BCs, permitindo uma transição suave e que demanda mais tempo no processo de realocação dos recursos (banda). No caso, a realocação de recursos \textit{soft} não impacta os usuários com recursos já alocados. Ambas as abordagens foram contempladas no BAMSDN e serão discutidas nas próximas seções.

\subsection{Explorando a criação de LSPs emuladas com o SDN OpenFlow}
A estratégia SDN/OpenFlow do BAMSDN tem dois papeis fundamentais no controle e programabilidade da rede: i) permite a criação de fluxos que emulam os LSPs existentes numa rede MPLS tradicional; e ii) permite uma migração suave de uma rede com conexões (LSPs) controladas pelo MPLS para uma rede com conexões (LSPs) com o controle baseado totalmente no OpenFlow.

O OpenFlow é utilizado no BAMSDN para a implantação do controle de operação dos \textit{switches} de forma logicamente centralizada e, assim sendo, suporta uma visão completa dos recursos e estado da rede para todos os seus roteadores e enlaces.

Uma vez que o módulo BAM do arcabouço libera a criação de um LSP, o OpenFlow é acionado para a criação de um fluxo que efetivamente emula o estabelecimento (\textit{setup}) de um LSP clássico do MPLS. Em outras palavras, um caminho comutado por software (\textit{SSP - Software Switched Path)} é criado através da configuração das tabelas de fluxo dos \textit{switches} na rota utilizada (\textit{datapath}) pelo OpenFlow. Assim sendo, o SSP equivale a um LSP MPLS clássico na visão externa da rede e, em termos da implementação BAMSDN, é um fluxo configurado que será referenciada como LSP MPLS neste artigo.

O BAMSDN migra a operação interna da rede para o paradigma SDN mas mantém a visão externa dos usuários no estilo MPLS. Este aspecto estratégico facilita a migração da operação interna da rede para um novo paradigma sem interferir diretamente na visão da rede comumente utilizada pelos usuários dos ISPs (\textit{Internet Service Providers}).

Um outro aspecto positivo da estratégia de utilização do OpenFlow no BAMSDN é o fato do BAM tratar da distribuição dos recursos. Em  efeito, quando uma operação de rede migra para o SDN/ OpenFlow, faz-se necessário o desenvolvimento de uma inteligência que aloca os recursos. No caso do BAMSDN, esta inteligência é provida pela operação dos BAM. Novamente, a solução permite uma migração suave para uma operação totalmente baseada no SDN/OpenFlow sem  a necessidade de um investimento significativo imediato na inteligência da operação da rede.

\subsection{Trabalhos relacionados}

Em anos recentes, alguns trabalhos têm discutido a utilização de BAMs para a alocação de recursos em diferentes tipos de rede.

Em \cite{hesselbach_management_2016} e \cite{reale_evaluating_2017} é proposto o uso de BAMs para a alocação de recursos em redes ópticas elásticas (\textit{Elastic Optical Network} - EON). Em \cite{goldberg_bandwidth_2006} e \cite{shan_bandwidth_2005} são avaliados parâmetros de operação de redes MPLS com BAM e \cite{trivisonno_network_2015} aplica modelo BAM para sistemas 5G numa estratégia VNE (\textit{Virtual Network Embedding}) e fazendo uso do SDN.

Em relação à utilização do SDN para o controle de operação de rede, o Google utiliza o OpenFlow para gerenciar e operar algumas de suas redes internas (\textit{backbone}
dos usuários e dos centro de dados) e indica um resultado positivo nesta implantação
com redução da complexidade do \textit{backbone} e de custos \cite{jain_b4:_2013}. A
solução Google (Rede B4), além da diferenciação pelo porte, tem uma estratégia diferente
da suportada pelo BAMSDN. Na solução Google não existe uma transição via o
MPLS e toda a inteligência para a computação de rotas, alocação de recursos, convergência após falhas e outros aspectos da operação da rede SDN/OpenFlow foram desenvolvidas pela empresa.

As vantagens de uma migração suave para o SDN é também discutida em \cite{tanha_traffic_2018} e é alinhada com a proposta do BAMSDN.

Alguns dos autores deste artigo apresentam em \cite{martins_g-bam:_2014,reale_preliminary_2014} resultados de simulação referente à utilização de BAMs (particularmente MAM, RDM e ATCS com o GBAM) para a otimização de recursos e numa perspectiva de aplicação para a gerência autonômica e cognitiva, porém sem a utilização das facilidades do paradigma SDN.

Importante ressaltar que nos trabalhos correlatos encontrados na literatura utiliza-se apenas configurações estáticas de BCs e para alguma eventual alteração destes parâmetros a rede deve ser reiniciada. Assim sendo, não foi identificada solução dinâmica com SDN semelhante à solução proposta.

Em resumo e, dentro do limite do nosso conhecimento, a literatura não trabalha a implantação de BAMs com o SDN/OpenFlow para uma rede MPLS visando a dinamicidade e a eficiência na alocação de banda numa perspectiva de migração de uma rede MPLS clássica para uma rede MPLS controlada pelo SDN/OpenFlow.

\section{Arcabouço BAMSDN}
\label{sec:bamsdn}

O arcabouço BAMSDN proposto neste trabalho visa oferecer suporte à implantação de modelos de alocação de banda (BAM) utilizando os benefícios e características do paradigma SDN. As próximas subseções detalham os componentes do arcabouço proposto assim como suas interações.

\begin{figure}[!t]
\centering
\includegraphics[width=1\linewidth]{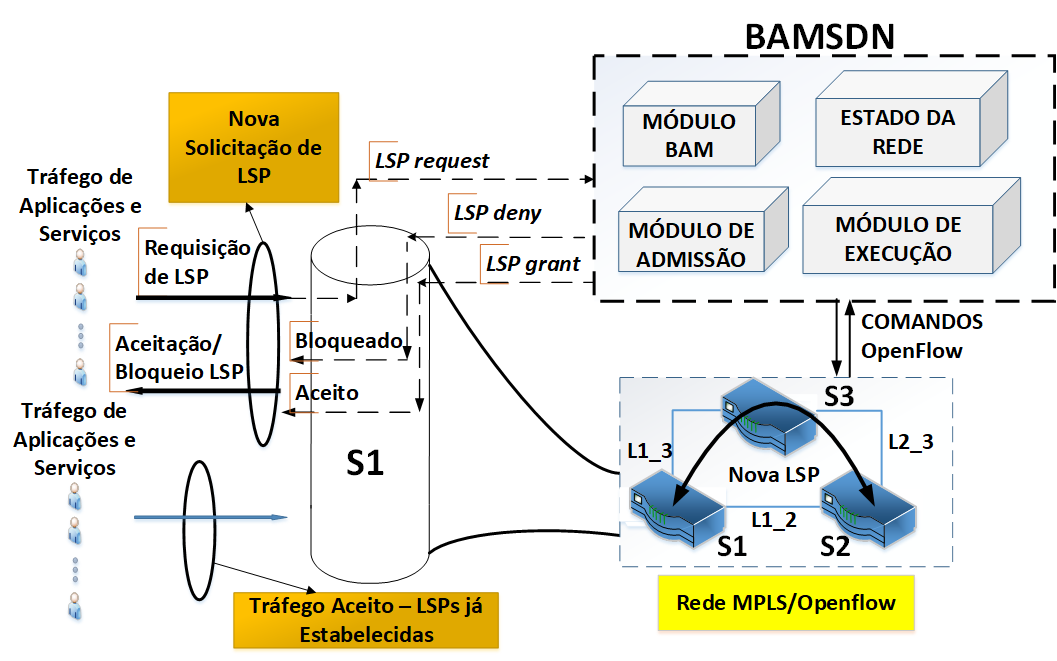}
\caption{Arcabouço BAMSDN e componentes básicos.}
\label{fig:bamVisaoGeralArcabouco}
\end{figure}

\subsection{Componentes do arcabouço BAMSDN}

O arcabouço BAMSDN é ilustrado na Figura~\ref{fig:bamVisaoGeralArcabouco}. Ele é implementado como um controlador SDN/OpenFlow com um conjunto de 4 módulos: (i) \textbf{Admissão}, responsável pelo controle das solicitações por recursos de rede (banda); (ii) \textbf{BAM}, que faz a verificação dos recursos, aplica o modelo de alocação de banda disponível e altera, quando necessário, as restrições de banda BCs (componente do estado da rede) visando a flexibilização no uso do recurso banda entre as classes; (iii) \textbf{Estado da Rede}, que mantém o estado da rede em termos dos LSPs estabelecidos, rotas utilizadas e utilização de banda por classe de tráfego; (iv) \textbf{Execução}, que efetiva a alocação do recurso banda através do estabelecimento, bloqueio, preempção (liberação) de LSPs utilizando o OpenFlow.

\begin{figure*}[!t]
\centering
\includegraphics[width=.6\textwidth]{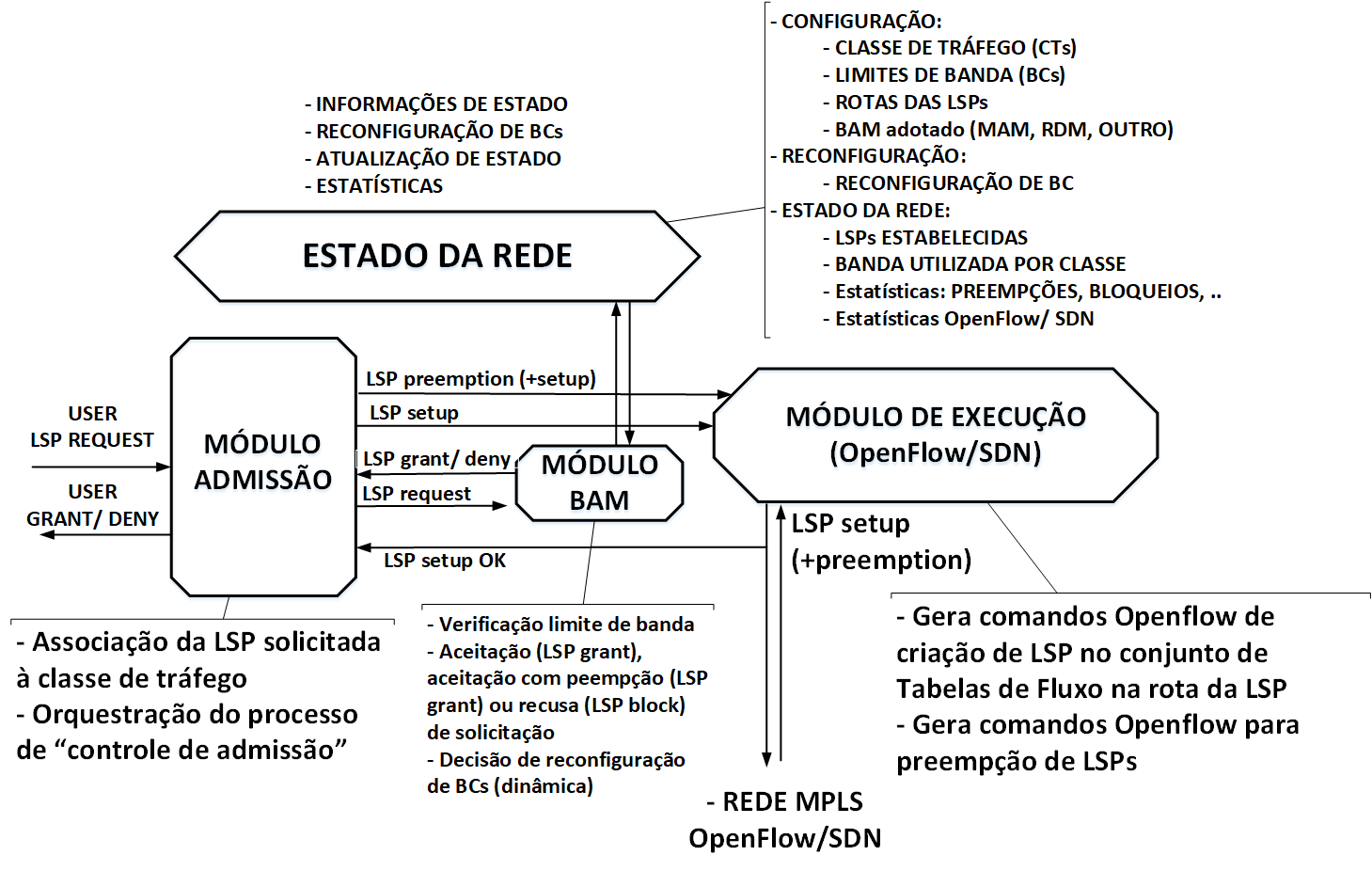}
\caption{Fluxo de admissão para requisição de LSPs.}
\label{fig:fluxoDeAdmissao}
\end{figure*}

De forma conjunta, tais módulos implementam o fluxo de admissão e controle de utilização dos recursos da rede (banda).

O BAMSDN suporta de forma simultânea a alocação de banda para os LSPs solicitados e a reconfiguração dinâmica da disponibilidade de banda (BC) por classe de aplicação agrupadas em classes de tráfego (CT).

\subsection{Fluxo de admissão}

A Figura~\ref{fig:fluxoDeAdmissao} ilustra o fluxo de admissão implementado pelo BAMSDN a partir de uma requisição de um LSP (LSP) usando um determinado quantitativo de banda. Neste fluxo são realizadas três ações principais: (i) controle de admissão, (ii) verificação de recursos; (iii) execução.

Quando um pacote associado a um fluxo de comunicação entre dois hospedeiros chega à rede, o \textit{switch} ao qual o hospedeiro de origem está conectado verifica se existe alguma informação correspondente ao fluxo na sua tabela de fluxos. Caso exista, o fluxo é encaminhado, uma vez que um LSP associado a este fluxo já foi criado. Caso contrário, um \textit{PacketIn} é encaminhado ao arcabouço BAMSDM atuando como controlador da rede para iniciar o processo de controle de admissão.

\subsubsection{Controle de admissão}

Na etapa de controle de admissão, o pacote reencaminhado é inicialmente processado pelo módulo de controle de admissão a partir de comparações (\textit{matching})\footnote{IP de origem, IP de destino, porta de origem e porta de destino entre outros} que permitem definir: i) A rota para o LSP a ser criado (associada a um fluxo); e ii) a classe de trafego do pacote. Tais informações são utilizadas para gerar uma nova instância de LSP e encaminhar a solicitação (\textit{LSP request}) para o módulo BAM.

O controle de admissão funciona como orquestrador do processo e, com a resposta do módulo BAM, dois conjuntos de ações são possíveis: i) o LSP é criado (\textit{LSP grant}); ou ii) o LSP é bloqueado (\textit{LSP deny}). Com a alocação de novo LSP aceito (\textit{LSP grant}) duas ações podem ser acionadas pelo controle de admissão: i) o LSP é criado imediatamente quando existem recursos suficientes; ii) o LSP é criado após a liberação de recursos (preempção de LSPs estabelecidos) na rede. Ambas as ações são solicitadas ao módulo de execução do BAMSDN. A execução bem sucedida da alocação de um novo LSP é retornada para a rede (\textit{user grant}) com o encaminhamento de pacotes para este fluxo sendo estabelecido pelo \textit{switches} envolvidos na trajetória (\textit{path}) do LSP. Com a rejeição para o estabelecimento de um novo LSP (\textit{LSP deny}), a rede é notificada e as estatísticas de bloqueio de LSPs atualizada no módulo de estado.
    
\subsubsection{Verificação de recursos}

A etapa de verificação de recursos é executada pelo módulo BAM do arcabouço. O modelo BAM em uso (p.ex., MAM, RDM ou outro), configurado na inicialização da rede, verifica a disponibilidade de recursos (banda) para todos os enlaces no caminho do LSP sendo requisitado. O processo acontece da seguinte forma: (i) Caso existam recursos disponíveis, a instância do novo LSP e a lista de portas logicas que o LSP utilizará em cada enlace são encaminhados para o Controle de Admissão; (ii) Caso não exista recurso disponível o modulo BAM identifica qual(is) recursos devem ser liberados (LSPs premptados ou devolvidos) e encaminha a informação para o Controle de Admissão.

\subsubsection{Execução}

Na etapa de execução, a instancia de um novo LSP e a mensagem de controle são processados, conforme a seguir: (i) Caso a mensagem de controle seja de `Bloqueio' (\textit{None}), a instância do LSP é descartado e uma mensagem (\textit{PacketOut}) de `Bloqueio' para o \textit{switch} de borda é enviada para fazer o descarte do fluxo (descarte de pacotes); (ii) Caso haja necessidade de interrupção de um ou mais LSPs é enviada uma mensagem (\textit{FlowMod}) para remoção destes LSPs da tabela de fluxo de todos os \textit{switches} em que eles estejam configurados. Por fim, caso a mensagem de controle seja aceitação do novo LSP, é enviada uma mensagem (\textit{FlowMod}) para cada \textit{switch} do caminho, que são configurados para estabelecer o limite largura de banda nas portas para o novo LSP.

\section{Experimentação}
\label{sec:experimentacao}

A verificação do BAMSDN foi realizada com uma avaliação experimental usando uma rede SDN/OpenFlow emulada. As próximas subseções descrevem a metodologia adotada, o ambiente de teste e os parâmetros de análise. 

\subsection{Metodologia}

Os objetivos principais dos experimentos realizados foram: i) avaliar o arcabouço BAMSDN com os modelos MAM e RDM para a alocação de banda; e ii) avaliar o OpenFlow realizando a emulação de LSPs e seus aspectos de migração.

Dois experimentos foram realizados. O primeiro experimento se baseia em LSPs geradas por aplicações de três classes de tráfego distintas, conforme a Tabela~\ref{tab-CTs_exp}. Neste experimento, foram utilizados os modelos de alocação de banda MAM e RDM, onde as solicitações de LSPs para as classes de trafego CT0 e CT1 ocorreram acima do limite estipulado para suas respectivas BCs (BC0 e BC1). As requisições dos LSPs ocorrem em 10 ciclos. Os LSPs da CT0 são iniciados no ciclo 0, os LSPs da CT1 são iniciados no ciclo 3 e CT2 no ciclo 6. Os parâmetros de emulação são: 10 ciclos de execução e 1.125 LSPs solicitados com 300s de tempo de vida cada. O critério de parada da emulação é de 1.125 LSPs solicitados.

\begin{table}[!t]
\renewcommand{\arraystretch}{1.3}
\caption{Tráfego Gerado por CT e Hospedeiros - Quantidade de LSPs.}
\label{tab-CTs_exp}
\centering
\begin{tabular}{ccccc}
\hline\hline
\multirow{2}{*}{\textbf{}} &
\multicolumn{4}{c}{\bfseries Experimento 1} \\ 

\bfseries Hospedeiro de Origem & \bfseries CT0 & \bfseries CT1 & \bfseries CT2 & \bfseries Total   \\ \hline
$H_{S1}$ & 400 & 70 & 0 & 470 \\
$H_{S2}$ & 400 & 70 & 0 & 470 \\
$H_{S3}$ & 200 & 35 & 20& 255 \\  \hline
Total & 1000 & 105 & 20 & 1125 \\  \hline
Ciclo de Início & 0 & 3 & 6 & - \\ \hline\hline

\multirow{2}{*}{\textbf{}} &
 \multicolumn{4}{c}{\bfseries Experimento 2} \\ 

\bfseries Hospedeiro de Origem & \bfseries CT0 & \bfseries CT1 & \bfseries CT2 & \bfseries Total  \\ \hline
$H_{S1}$ & 300 & 90 & 0 & 390\\
$H_{S2}$ & 200 & 90 & 0 &290\\
$H_{S3}$ & 0 & 45 & 45 &90\\  \hline
Total & 500 & 225 & 45 & 770\\  \hline
Ciclo de Início & 0 & 1 & 1& -\\ \hline\hline

\end{tabular}
\end{table}

O segundo experimento tem como objetivo avaliar a reconfiguração de BCs, utilizando o MAM, com as estratégias \textit{hard} e \textit{soft} discutidas em \cite{reale_preliminary_2014} e apresentadas na seção \ref{Problema}. Foi utilizado um tráfego de entrada de solicitação de LSPs conforme apresentado na Tabela \ref{tab-CTs_exp}. As solicitações de tais LSPs, para as classes de trafego CT0 e CT1, alteraram suas configurações de limite de restrição de banda em tempo de execução com o objetivo da classe CT1 ter uma restrição de banda maior que a classe CT0. As requisições dos LSPs ocorrem em 10 ciclos. Os LSPs da CT0 são iniciados no ciclo 0 e as requisições dos LSPs de CT1  e CT2 são iniciadas no ciclo 1. Os parâmetros de emulação são: 10 ciclos de execução e 770 LSPs com 300s de tempo de vida cada. O critério de parada é de 770 LSPs solicitados.   

\subsection{Ambiente de experimentação}

Os experimentos seguiram uma abordagem de emulação que reproduz uma rede OpenFlow usando o \textit{Mininet}\footnote{http://mininet.org/}. Para isso, foram utilizados: (i)~\textit{PC Core} i5, 2.9\textit{Ghz}, 8\textit{GB} de \textit{RAM}; (ii)~Sistema Operacional \textit{Ubuntu Server 15.04}, \textit{x}64, versão do \textit{kernel} 3.19.0-30; (iii) \textit{Mininet}, versão 1.8r11 customizado; (iv) Protocolo \textit{OpenFlow}\footnote{http://archive.openflow.org/}, versão 1.0; (vii)~Controlador \textit{OpenFlow} POX\footnote{https://github.com/noxrepo/pox}, versão 0.2.0; e (viii)~Gerador de tráfego iPerf3, versão 3.0.7.

\begin{figure}[!t]
\centering
\includegraphics[width=1\linewidth]{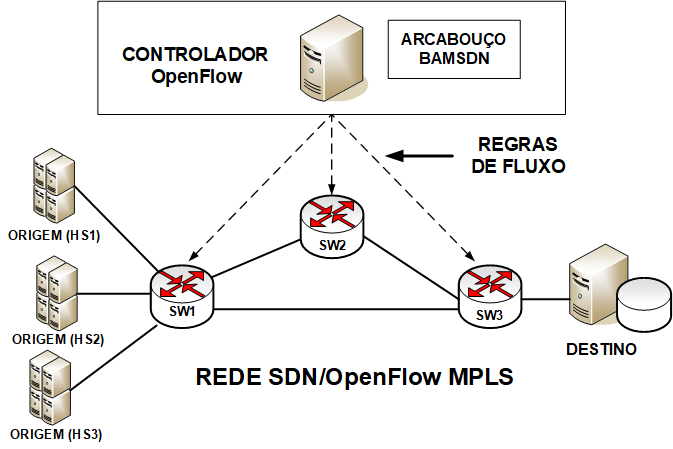}
\caption{Rede MPLS OpenFlow emulada utilizada nos experimentos.}
\label{fig:topologiaValidacao}
\end{figure}

O ambiente de rede emulado nos experimentos é ilustrado na Figura~\ref{fig:topologiaValidacao}. O mesmo é constituído de três hospedeiros, que atuam como origem de tráfego ($H_{S1}$, $H_{S2}$ e $H_{S3}$), um hospedeiro que atua como destino, um controlador OpenFlow que configura as regras de fluxos de acordo com as instruções do BAMSDN e, por fim, três \textit{switches} que fazem a conexão entre os hospedeiros de origem e destino. 

\subsection{Parâmetros da experimentação}

Os parâmetros globais para o cenário de experimentação são os seguintes: (i) enlaces com capacidades de 500 Mbps; ii) três classes de tráfego: CT0, CT1 e CT2, com larguras de banda máxima de LSP de 5 Mbps, 10 Mbps e 20 Mbps, respectivamente; e iii) restrições de largura de banda (\textit{Bandwidth  Constraint} (BC)), de acordo com a Tabela~\ref{tab-bcPorCct}.

\begin{table}[!t]
\renewcommand{\arraystretch}{1.3}
\caption{Restrições de Banda (BC) por Classes de Tráfego (CT)}
\label{tab-bcPorCct}
\centering
\begin{tabular}{cccc}
\hline \hline
\multirow{2}{*}{\textbf{}} &
\multicolumn{3}{c}{\textbf{MAM}} \\ & \textbf{Max (\%)}	& \textbf{Max (Mbps)} & \textbf{CTs}   \\ \hline
BC0 & 50 & 250 & CT0  \\
BC1 & 30 & 150 & CT1 \\
BC2 & 20 & 100 & CT2 \\  \hline \hline

\multirow{2}{*}{\textbf{}} &
 \multicolumn{3}{c}{\textbf{RDM}} \\ 

 & \textbf{Max (\%)} &  \textbf{Max (Mbps)} & \textbf{CTs}  \\ \hline
BC0 & 100& 500 & CT0 + CT1 + CT2 \\
BC1 & 50 & 250 & CT1 + CT2\\
BC2 & 20 & 100 & CT2 \\  \hline\hline

\end{tabular}
\end{table}

\section {Análise dos resultados}
\label{sec:analise}
 
Os resultados obtidos na alocação de banda pelo BAMSDN utilizando o mecanismo BAM com SDN são analisados através da verificação dos parâmetros de utilização, preempção e bloqueio para os modelos MAM e RDM com e sem a reconfiguração de BC. A experimentação está disponível no GitHub (https://github.com/EliseuTorres/BAMSDN).

\subsection{Resultados do BAMSDN com os modelos MAM e RDM}\label{AnaliseA}

\begin{figure*}[!t]
\centering
\subfloat[Utilização e bloqueio com o MAM.]{\includegraphics[width=2.5in]{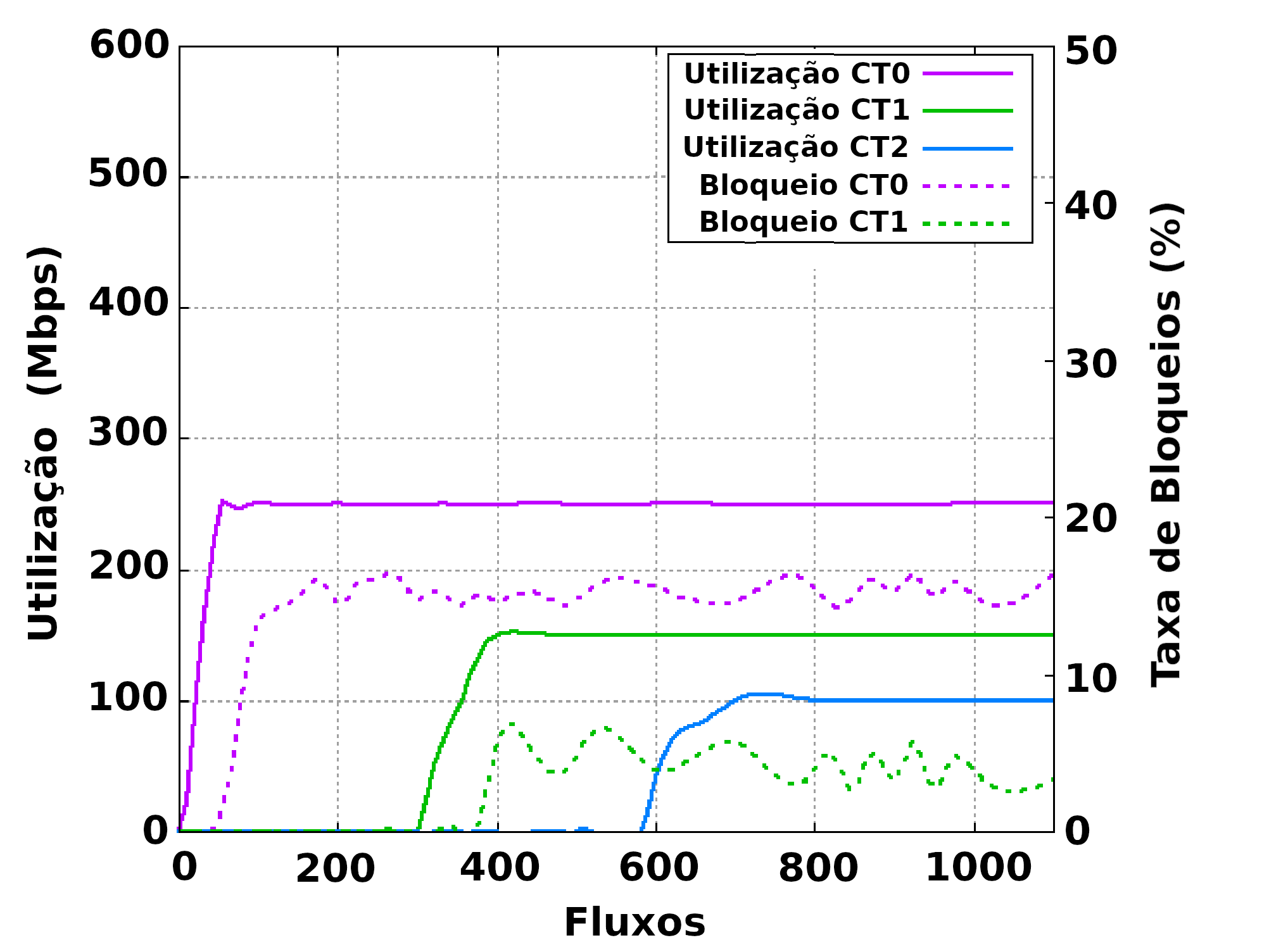}%
\label{fig:MAM}}
\hfil
\subfloat[Utilização e bloqueio com o RDM.]{\includegraphics[width=2.5in]{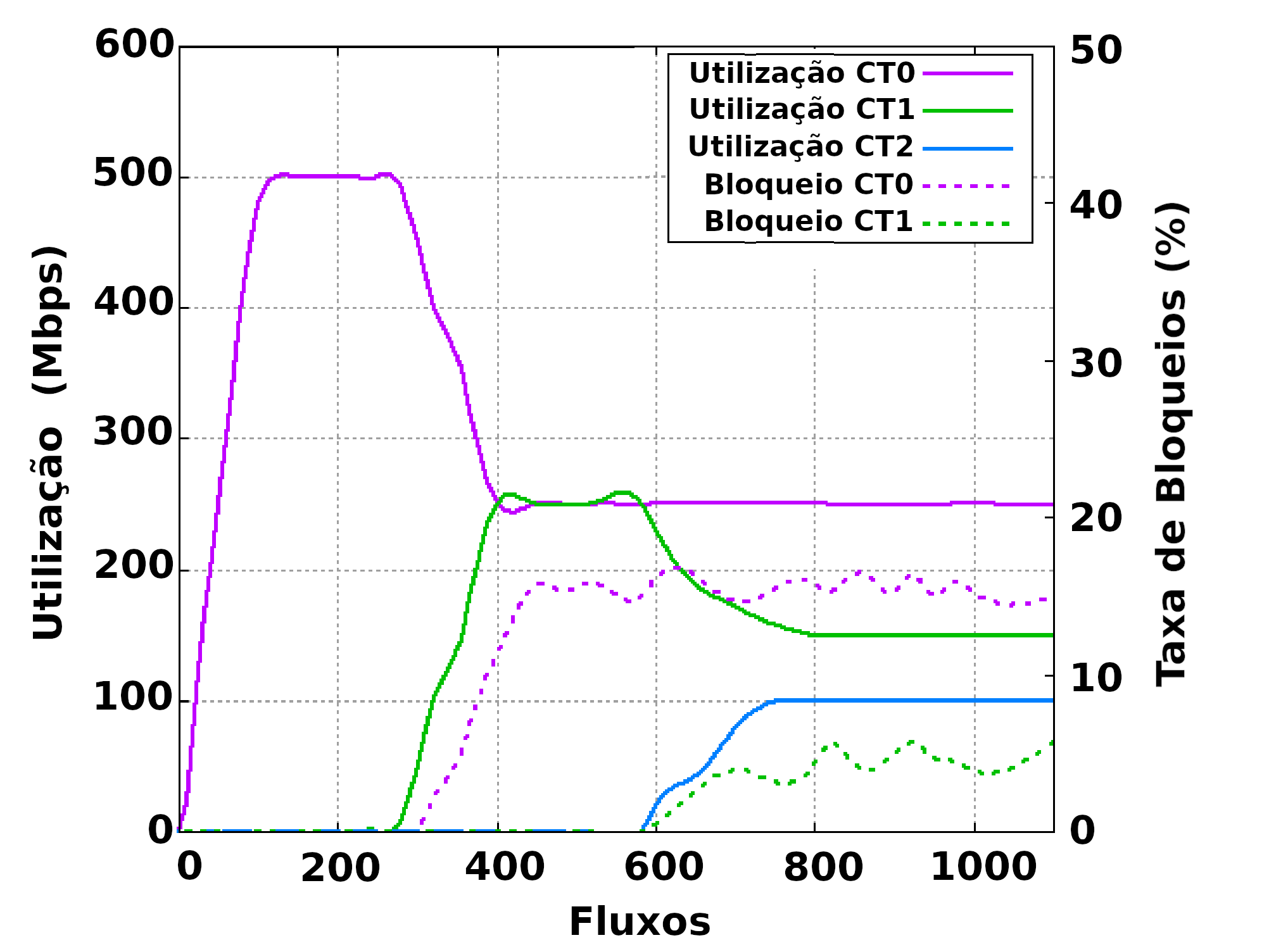}%
\label{fig:RDM}}
\caption{Utilização e taxa de bloqueio do BAMSDN usando os modelos MAM e RDM.}
\label{fig:MAMRDM}
\end{figure*}

A Figura~\ref{fig:MAMRDM} apresenta os resultados obtidos pelo BAMSDN para a utilização e as taxas de bloqueio com a utilização dos modelos MAM (Fig. ~\ref{fig:MAMRDM}a) e RDM (Fig. ~\ref{fig:MAMRDM}b).

A Figura~\ref{fig:MAM} apresenta a utilização e a taxa de bloqueio por CT para o modelo MAM e ilustra a baixa eficiência deste modelo. No MAM os recursos de cada classe são privados e não podem ser compartilhados por outras classes. Assim sendo, a adoção do MAM pelo BAMSDN leva a uma menor eficiência em relação à utilização da banda total disponível para o enlace. Em efeito, usando o MAM, o enlace fica com uma banda ociosa de 250 Mbps entre os fluxos 50 e 300 já ocorrendo bloqueios da ordem de 15\% para CT0. Entre os fluxos 400 e 600 tem-se uma banda ociosa de 150 Mbps e ocorrem bloqueios para CT0 da ordem de 20\% e para CT1 da ordem de 5\%. O número de LSPs atendidos correspondem ao limite permitido pela BC configurada para cada CT: (250Mbps - CT0, 150Mbps - CT1 e 100Mbps - CT2). Com os limites de BC atingidos, a utilização do enlace se mantém constante com bloqueios acontecendo para a parte da demanda de LSPs que não pode ser atendida pela liberação de banda resultante do encerramento dos LSPs ativos.

A Figura~\ref{fig:RDM} apresenta a utilização e a taxa de bloqueio por CT com a adoção do modelo RDM. Isso resulta no BAMSDN apresentando uma melhor eficiência na utilização do enlace e redução do bloqueio em relação ao mesmo cenário de tráfego da avaliação anterior usando o MAM. No RDM, CTs de classe inferior podem usar a banda disponível nos CTs de classe superiores e isso é verificado na Figura~\ref{fig:RDM} entre os fluxos 0 e 300 (ciclos 1 e 2) quando CT0 chega a utilizar a totalidade da banda disponível no enlace (500 Mbps). No BAMSDN com RDM classes que obtiveram banda de outras classes devolvem o recurso quando solicitado pela classe que o emprestou, resultando em preempções de LSPs já estabelecidas. Este aspecto da operação do BAMSDN com RDM é verificado na Figura~\ref{fig:RDM} entre os fluxos 300 e 400 quando LSPs da CT1 começam a ser solicitadas ao BAMSDN. As preempções são observadas na Figura~\ref{fig:RDM} indiretamente pela redução da utilização do enlace pela classe CT0 (de 500 para 250 Mbps). No BAMSDN com RDM os bloqueios acontecem quando as CTs passam a utilizar a sua banda alocada. Na Figura~\ref{fig:RDM} isso acontece entre os fluxo 300 e 1200 com uma taxa de bloqueio da ordem de 15\% para CT0 e em torno de 5\% para CT1. De maneira geral, observa-se comparando as Figuras~\ref{fig:RDM} e \ref{fig:MAM} que o RDM gerou um número inferior de bloqueios em relação ao MAM e gerou uma taxa de 6,12\% de preempções em CT0 e 11,63\% de preempções em CT1.

\subsection{Resultados do BAMSDN com a reconfiguração de BC}

A reconfiguração de restrição de banda (BC) pelo BAMSDN permite um nível adicional opcional de melhoria da eficiência da rede, mesmo para um modelo rígido como o MAM que, conforme apresentado na seção \ref{AnaliseA}, não permite uma otimização da utilização da banda disponível do enlace.

\begin{figure*}[!t]
\centering
\subfloat[Abordagem \textit{Hard}.]{\includegraphics[width=2.5in]{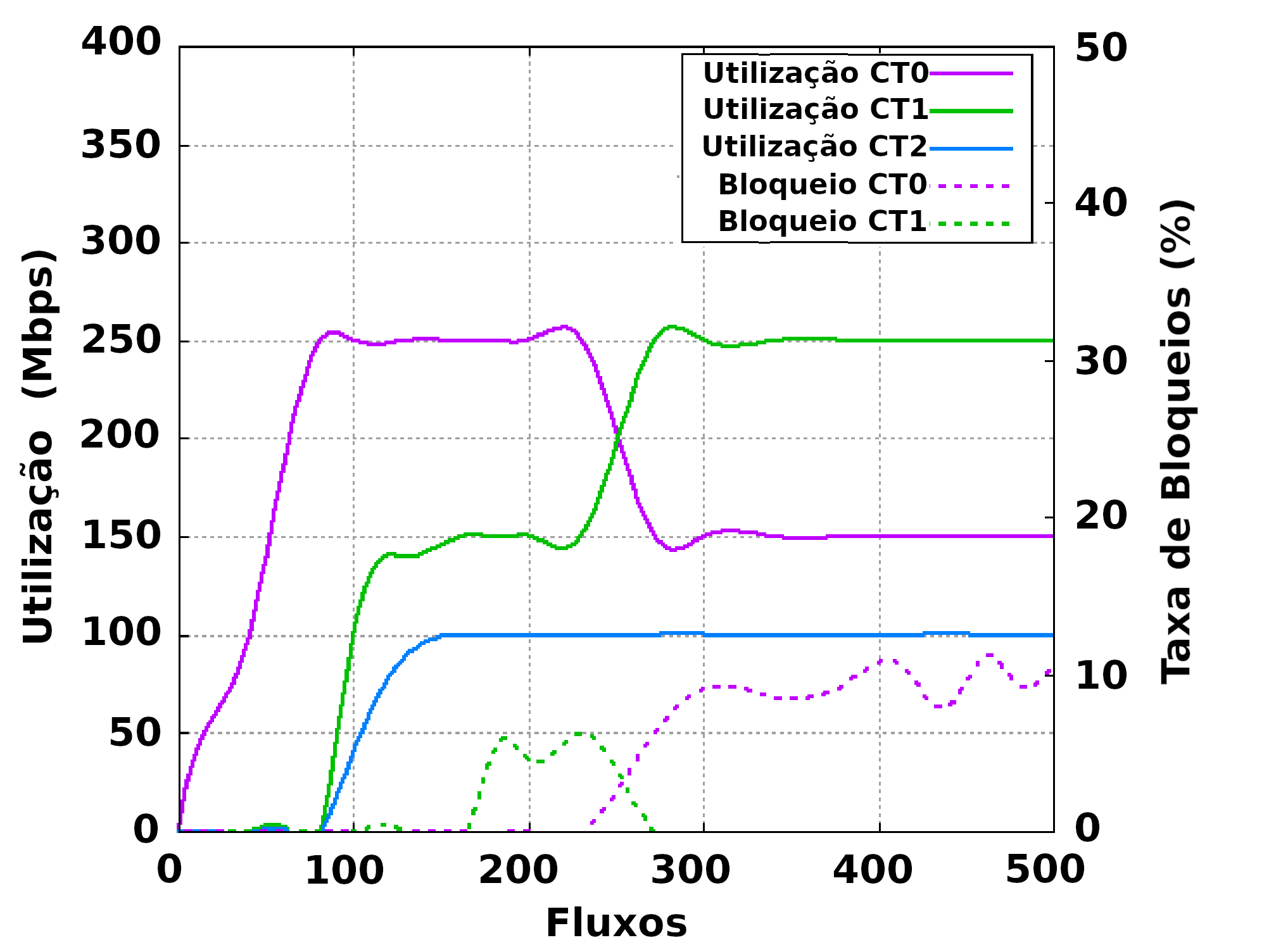}%
\label{fig:exp2UtilizacaoA}}
\hfil
\subfloat[Abordagem \textit{Soft}.]{\includegraphics[width=2.5in]{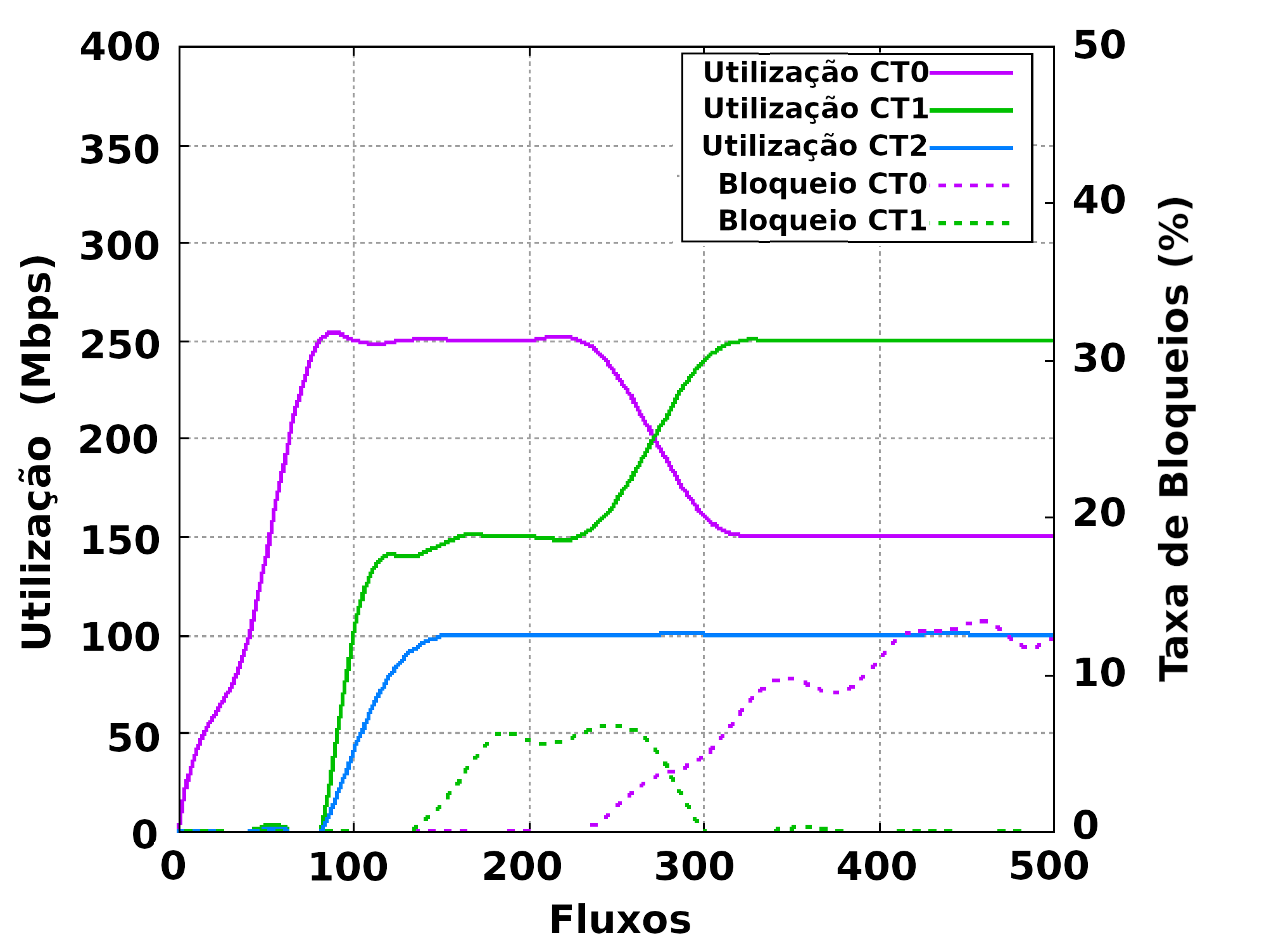}%
\label{fig:exp2UtilizacaoB}}
\caption{Reconfiguração de restrição de banda (BC) de CT com abordagens \textit{hard} e \textit{soft}.}
\label{fig:exp2Utilizacao}
\end{figure*}

As Figuras~\ref{fig:exp2UtilizacaoA} e~\ref{fig:exp2UtilizacaoB} apresentam a utilização e a taxa de bloqueio por CT resultantes  da reconfiguração de BC realizada pelo BAMSDN com as abordagens \textit{hard} e \textit{soft} respectivamente (Seção \ref{Problema}). O modelo adotado é o MAM utilizando os BCs definidos na Tabela \ref{tab-bcPorCct}. Importante ressaltar que a decisão de alterar os BCs foi introduzida na sequência de testes de forma a demonstrar que esta abordagem permite uma melhora na utilização e na taxa de bloqueio. Na operação de uma rede usando o BAMSDN, a decisão de alterar o BC e os valores da alteração devem ser computados dinamicamente utilizando critério ou política de operação da rede.

A Figura~\ref{fig:exp2UtilizacaoA} apresenta os resultados do experimento com a abordagem \textit{hard}. As classes CT0, CT1 e CT2 são saturadas entre os fluxos 0 e 300 com LSPs requisitadas para CT1 com bloqueio de até 5\%. No ciclo entre os fluxos 200 e 300 ocorre a reconfiguração \textit{hard} de BC1 para 150 Mbps e BC2 para 100 Mbps. O resultado é que o bloqueio de CT2 cai para 0\% e o bloqueio de CT0 cresce até 10\% pois houve uma redução da banda disponível para a classe. Do ponto de vista da operação da rede, tal opção de reconfiguração de BC pode subsidiar, por exemplo, uma opção da gerência da rede em alocar dinamicamente mais banda para uma classe mais prioritária que outra na medida em que existe uma demanda de LSPs a serem atendidas para a classe mais prioritária. Na experimentação realizada a classe CT0 teve uma taxa de preempção de 3,68\% após a reconfiguração de BCs. Isso representa uma espécie de custo de operação imposto à classe CT0 que é menos prioritária.

A Figura~\ref{fig:exp2UtilizacaoB} ilustra os resultados com a utilização da abordagem \textit{soft}. Observa-se que a efetivação da reconfiguração de BC fica mais alongada pelo fato da liberação e realocação dos recursos ocorrer apenas após o encerramento normal das LSPs ativas. Em outras palavras, não se força nenhuma preempção. Assim sendo, CT1 e CT2 têm um maior número absoluto de bloqueios em relação à opção \textit{hard}.

Por último, cumpre indicar que os resultados obtidos na experimentação com o MiniNet coincidem com as simulações realizadas nos trabalhos relacionados \cite{martins_g-bam:_2014} e \cite{reale_alloctc-sharing:_2011} para os modelos MAM e RDM. Isso sinaliza que, apesar do MiniNet ser um mecanismo controlado com algumas limitações de escala, isso não impactou na qualidade dos dados obtidos na validação do BAMSDN.

\subsection{Vantagens da utilização do SDN/OpenFlow no BAMSDN}

A adoção do SDN/ OpenFlow na implementação do arcabouço BAMSDN apresenta benefícios em relação à uma rede MPLS/DS-TE (\textit{DiffServ Aware Traffic Engineering}) clássica. Estes benefícios são de dois tipos: i) redução da capacidade computacional e da complexidade de configuração da malha de roteadores da rede; e ii)  simplificação da estrutura de monitoramento da rede visando a coleta de dados de operação para a alocação de banda.

A redução da capacidade computacional e da complexidade de configuração da malha de roteadores é obtida através da centralização no BAMSDN do cálculo de rotas para o estabelecimento de novas LSPs, da centralização da execução do procedimento de preempção de LSPs estabelecidas e da centralização do cálculo de disponibilidade de banda por enlace. Em termos práticos, isso significa que os roteadores utilizados com o BAMSDN não necessitam de nenhum protocolo associado à operação do DS-TE, tipo o RSVP-TE (\textit{Resource Reservation Protocol with Traffic Engineering}) ou o  OSPF (\textit{Open Shortest Path First}), podendo ser implementados com \textit{switches} no estilo do SDN. Isso representa um ganho significativo em relação à implantação (\textit{deployment}) e operação da rede.

A simplificação do monitoramento é obtida através da execução centralizada do OpenFlow numa máquina controladora (BAMSDN). O OpenFlow do BAMSDN coleta e centraliza os dados de operação dos roteadores permitindo uma visão tipo \textit{snapshot} de todo o conjunto de rotedores utilizados. Além disso, o BAMSDN mantém o estado da rede em termos dos LSPs estabelecidos, da utilização de banda por CT e da configuração do BAM para os enlaces.

\subsection{Considerações em relação à escalabilidade do BAMSDN}

A escalabilidade do arcabouço BAMSDN depende principalmente de como o BAM é utilizado na alocação de banda para todos os enlaces da rede MPLS.

O modelo BAM controla a alocação banda por enlace de forma independente e a alocação de uma nova LSP requer a verificação da disponibilidade de banda para todos os enlaces no caminho (\textit{path}). O BAMSDN centraliza todas as informações de banda por CT na sua base de dados. Assim sendo, entende-se que o BAMSDN escala para a alocação de banda numa rede com ``N" roteadores considerando que, tipicamente, redes grandes têm sempre um número relativamente pequeno (N\textless40) de roteadores por trajetória e que o custo computacional do BAM é o acesso de leitura à base de dados do BAMSDN.

\section{Considerações Finais e Trabalhos Futuros}
\label{sec:conclusao}

Em relação às soluções com BAM existentes, os ganhos principais do BAMSDN são a melhor utilização da banda dos enlaces e a redução dos bloqueios com o RDM. Além disso, a reconfiguração de BCs propiciada pelo BAMSDN permite uma adequação dinâmica da alocação de banda ao perfil de tráfego de entrada. Isso tanto em relação ao comportamento como em relação à configuração do MAM e do RDM.

O arcabouço BAMSDN, em resumo, agrupa as vantagens da utilização do MAM e do RDM para a alocação de banda com a utilização do OpenFlow para o controle da rede MPLS. Este é um resultado inovador em relação às soluções de alocação de recursos com BAM existentes. Em efeito, o BAMSDN permite a redução da capacidade computacional e simplificação da configuração dos roteadores da rede, permite a monitoração centralizada de toda a rede e, por último, facilita a migração de uma rede MPLS clássica para uma rede MPLS programada pelo OpenFlow.

Em termos de trabalho futuro, pretende-se implantar o arcabouço BAMSDN na Rede de Serviços FIBRE\footnote{fibre.org.br} e na rede experimental BAMBU\footnote{https://www.pop-ba.rnp.br/Bambu}, visando avaliar a operação do BAMSDN com programação SDN/OpenFlow distribuída.

\ifCLASSOPTIONcaptionsoff
  \newpage
\fi

\bibliographystyle{IEEEtran}
\bibliography{IEEEabrv,sbc-template.bib}

\begin{thebibliography}{10}
\providecommand{\url}[1]{#1}
\csname url@samestyle\endcsname
\providecommand{\newblock}{\relax}
\providecommand{\bibinfo}[2]{#2}
\providecommand{\BIBentrySTDinterwordspacing}{\spaceskip=0pt\relax}
\providecommand{\BIBentryALTinterwordstretchfactor}{4}
\providecommand{\BIBentryALTinterwordspacing}{\spaceskip=\fontdimen2\font plus
\BIBentryALTinterwordstretchfactor\fontdimen3\font minus
  \fontdimen4\font\relax}
\providecommand{\BIBforeignlanguage}[2]{{%
\expandafter\ifx\csname l@#1\endcsname\relax
\typeout{** WARNING: IEEEtran.bst: No hyphenation pattern has been}%
\typeout{** loaded for the language `#1'. Using the pattern for}%
\typeout{** the default language instead.}%
\else
\language=\csname l@#1\endcsname
\fi
#2}}
\providecommand{\BIBdecl}{\relax}
\BIBdecl

\bibitem{rachkidi_cloud_2017}
E.~Rachkidi, D.~Belaïd, N.~Agoulmine, and N.~Chendeb,
  ``\BIBforeignlanguage{en}{Cloud of {Things} {Modeling} for {Efficient} and
  {Coordinated} {Resources} {Provisioning}},'' in
  \emph{\BIBforeignlanguage{en}{{OTM} {Confederated} {Int.} {Conf.}}}\hskip 1em
  plus 0.5em minus 0.4em\relax Springer, Sep. 2017, pp. 175--193.

\bibitem{zuo_fast_2015}
L.~Zuo, M.~Zhu, and Q.~Wu, ``\BIBforeignlanguage{en}{Fast and {Efficient}
  {Bandwidth} {Reservation} {Algorithms} for {Dynamic} {Network}
  {Provisioning}},'' \emph{\BIBforeignlanguage{en}{Journal of Network and Syst.
  Manag.}}, vol.~23, no.~3, pp. 420--444, Jul. 2015.

\bibitem{fan_dynamic_2016}
B.~Fan, S.~Leng, and K.~Yang, ``A {Dynamic} {Bandwidth} {Allocation}
  {Algorithm} in {Mobile} {Networks} with {Big} {Data} of {Users} and
  {Networks},'' \emph{IEEE Network}, vol.~30, no.~1, pp. 6--10, Jan. 2016.

\bibitem{martins_uma_2015}
J.~Martins, R.~Bezerra, G.~Durães, and R.~Reale,
  ``\BIBforeignlanguage{Portuguese}{Uma {Visão} {Tutorial} dos {Modelos} de
  {Alocação} de {Banda} como {Mecanismo} de {Provisionamento} de {Recursos}
  em {Redes} {IP}/{MPLS}},'' \emph{\BIBforeignlanguage{Portuguese}{Revista de
  Sistemas e Computação}}, vol.~5, no.~2, pp. 144--155, Dec. 2015.

\bibitem{trivisonno_network_2015}
R.~Trivisonno, R.~Guerzoni, I.~Vaishnavi, and A.~Frimpong, ``Network {Resource}
  {Management} and {QoS} in {SDN}-{Enabled} 5g {Systems},'' in \emph{{IEEE}
  {GLOBECOM} 2015}.\hskip 1em plus 0.5em minus 0.4em\relax IEEE, Dec. 2015, pp.
  1--7.

\bibitem{pistek_-mar:_2015}
M.~Pištek, M.~Medvecký, and S.~Klučik, ``A-{MAR}: {A} {New} {Bandwidth}
  {Constraint} {Model} for {DS}-{TE} {Networks},'' in \emph{38th {Int.} {Conf.}
  on {Telecommunications} and {Signal} {Processing}}, Jul. 2015, pp. 1--5.

\bibitem{onali_traffic_2008}
T.~Onali and L.~Atzori, ``Traffic {Engineering} in {Next} {Generation}
  {Networks} {Using} {Genetic} {Algorithms},'' in \emph{{GLOBECOM}}, Nov. 2008,
  pp. 1--5.

\bibitem{kreutz_software-defined_2014}
D.~Kreutz, F.~Ramos, P.~Verissimo, C.~Rothenberg, S.~Azodolmolky, and S.~Uhlig,
  ``Software-{Defined} {Networking}: {A} {Comprehensive} {Survey},''
  \emph{Proc. of IEEE}, vol. 103, no.~1, pp. 14--76, Dec. 2014.

\bibitem{faucher_maximum_2005}
F.~L. Faucher and W.~Lai, ``Maximum {Allocation} {Bandwidth} {Constraints}
  {Model} for {DSTE},'' Tech. Rep. RFC 4125, Jun. 2005.

\bibitem{pinto_da_costa_neto_algoritmos_2008}
W.~Neto, S.~Brito, and J.~Martins, ``\BIBforeignlanguage{Portuguese}{Algoritmos
  de {Seleção} de {Caminho} e {Gerenciamento} de {Banda} {Compartilhada}
  conforme ao {Modelo} {RDM} para {Classes} de {Tráfego} em {Rede}
  {DS}-{TE}},'' in \emph{\BIBforeignlanguage{Portuguese}{{SBRC} 2008}}, Rio de
  Janeiro, May 2008, pp. 537--552.

\bibitem{reale_alloctc-sharing:_2011}
R.~F. Reale, W.~d. C.~P. Neto, and J.~S.~B. Martins, ``{AllocTC}-sharing: {A}
  {New} {Bandwidth} {Allocation} {Model} for {DS}-{TE} {Networks},'' in
  \emph{{LANOMS} 2011}.\hskip 1em plus 0.5em minus 0.4em\relax Equador: IEEE,
  Oct. 2011, pp. 1--4.

\bibitem{martins_g-bam:_2014}
J.~Martins, R.~Reale, and R.~Bezerra, ``G-{BAM}: {A} {Generalized} {Bandwidth}
  {Allocation} {Model} for {IP}/{MPLS}/{DS}-{TE} {Networks},'' \emph{Int.
  Journal of Computer Information Systems and Industrial Management Applic.},
  vol.~6, pp. 635--643, Dec. 2014.

\bibitem{reale_preliminary_2014}
R.~Reale, R.~Bezerra, and J.~Martins, ``\BIBforeignlanguage{English}{A
  {Preliminary} {Evaluation} of {Bandwidth} {Allocation} {Model} {Dynamic}
  {Switching}},'' \emph{\BIBforeignlanguage{English}{Int. Journal of Computer
  Networks \& Comm.}}, vol.~6, no.~3, pp. 131--143, May 2014.

\bibitem{hesselbach_management_2016}
X.~Hesselbach, J.~Amazonas, and J.~Botero, ``Management of resources under
  priorities in {EON} using a modified {RDM} based on the squatting-kicking
  approach,'' in \emph{18th {Int.} {Conf.} on {Transparent} {Optical}
  {Networks}}, 2016, pp. 1--5.

\bibitem{reale_evaluating_2017}
R.~Reale, R.~Bezerra, G.~Durães, A.~Soares, and J.~Martins,
  ``\BIBforeignlanguage{en}{Evaluating the {Applicability} of {Bandwidth}
  {Allocation} {Models} for {EON} {Slot} {Allocation}},'' in
  \emph{\BIBforeignlanguage{en}{{IEEEANTS} 2017}}, Dec. 2017, pp. 1--6.

\bibitem{goldberg_bandwidth_2006}
J.~B. Goldberg, S.~Dasgupta, and J.~C.~d. Oliveira, ``Bandwidth {Constraint}
  {Models}: {A} {Performance} {Study} with {Preemption} on {Link} {Failures},''
  in \emph{{IEEE} {Globecom} 2006}, Nov. 2006, pp. 1--5.

\bibitem{shan_bandwidth_2005}
T.~Shan and O.~W.~W. Yang, ``Bandwidth {Preemption} {Algorithms} for
  {Differentiated} {Service} aware {Traffic} {Engineering},'' in
  \emph{{GLOBECOM} 2005}, vol.~1, Nov. 2005, pp. 535--539.

\bibitem{jain_b4:_2013}
S.~Jain, A.~Kumar, S.~Mandal, J.~Ong, L.~Poutievski, A.~Singh, S.~Venkata,
  J.~Wanderer, J.~Zhou, M.~Zhu, J.~Zolla, U.~Hölzle, S.~Stuart, and A.~Vahdat,
  ``B4: {Experience} with a {Globally}-deployed {Software} {Defined} {Wan},''
  in \emph{{ACM} {SIGCOMM} 2013}, USA, 2013, pp. 3--14.

\bibitem{tanha_traffic_2018}
M.~Tanha, D.~Sajjadi, R.~Ruby, and J.~Pan, ``Traffic {Engineering}
  {Enhancement} by {Progressive} {Migration} to {SDN},'' \emph{IEEE
  Communications Letters}, vol.~22, no.~3, pp. 438--441, Mar. 2018.

\end{thebibliography}

\begin{IEEEbiography}[{\includegraphics[width=1in,height=1.25in,clip,keepaspectratio]{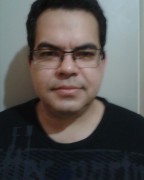}}]{MSc. Eliseu Silva Torres -}
Bachelor in Computer Science from Salvador University - UNIFACS (2014). He got his masters at the Graduate Program in Computer Science (PGCOMP) of Universidade Federal da Bahia (UFBA), where he participates in the research group Infrastructure and Systems for Networks and Telecommunications (INSERT). He is also a member of BAMBU, a metropolitan network of experimentation and innovation in the Future Internet.
\end{IEEEbiography}

\begin{IEEEbiography}[{\includegraphics[width=1in,height=1.25in,clip,keepaspectratio]{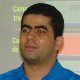}}]{Prof. Dr. Rafael F. Reale -}
Professor at Institute Federal of Bahia (IFBA). PhD. in Computer Science with DMCC (UFBA/UNIFACS/UEFS), MSc. in Computer Systems by Salvador University - UNIFACS (2011) and bachelor in Informatics by Universidade Católica do Salvador (2005). Professor at Instituto Federal
da Bahia (IFBA). His current research interests include Bandwidth Allocation Model, MPLS, DS-TE, Autonomy, QoS, Future Internet Architectures, Software-Defined Networks and Cognitive Networks.
\end{IEEEbiography}

\begin{IEEEbiography}[{\includegraphics[width=1in,height=1.25in,clip,keepaspectratio]{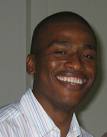}}]{Prof. Dr. Leobino N. Sampaio -}
Associate Professor of Computer Science at Federal University of Bahia (UFBA) and Member of the Brazilian Computer Society (SBC). He holds Ph.D. degree in Computer Science at Federal University of Pernambuco (UFPE) in 2011 and holds B.Sc. and M.Sc. degrees in Computer Science, both at Salvador University (UNIFACS) in 1996 and 2002, respectively. His current research interests include Future Internet Architectures, Software-Defined Networking, Information-Centric Networks, and Network Performance Evaluation. 
\end{IEEEbiography}

\begin{IEEEbiography}[{\includegraphics[width=1in,height=1.25in,clip,keepaspectratio]{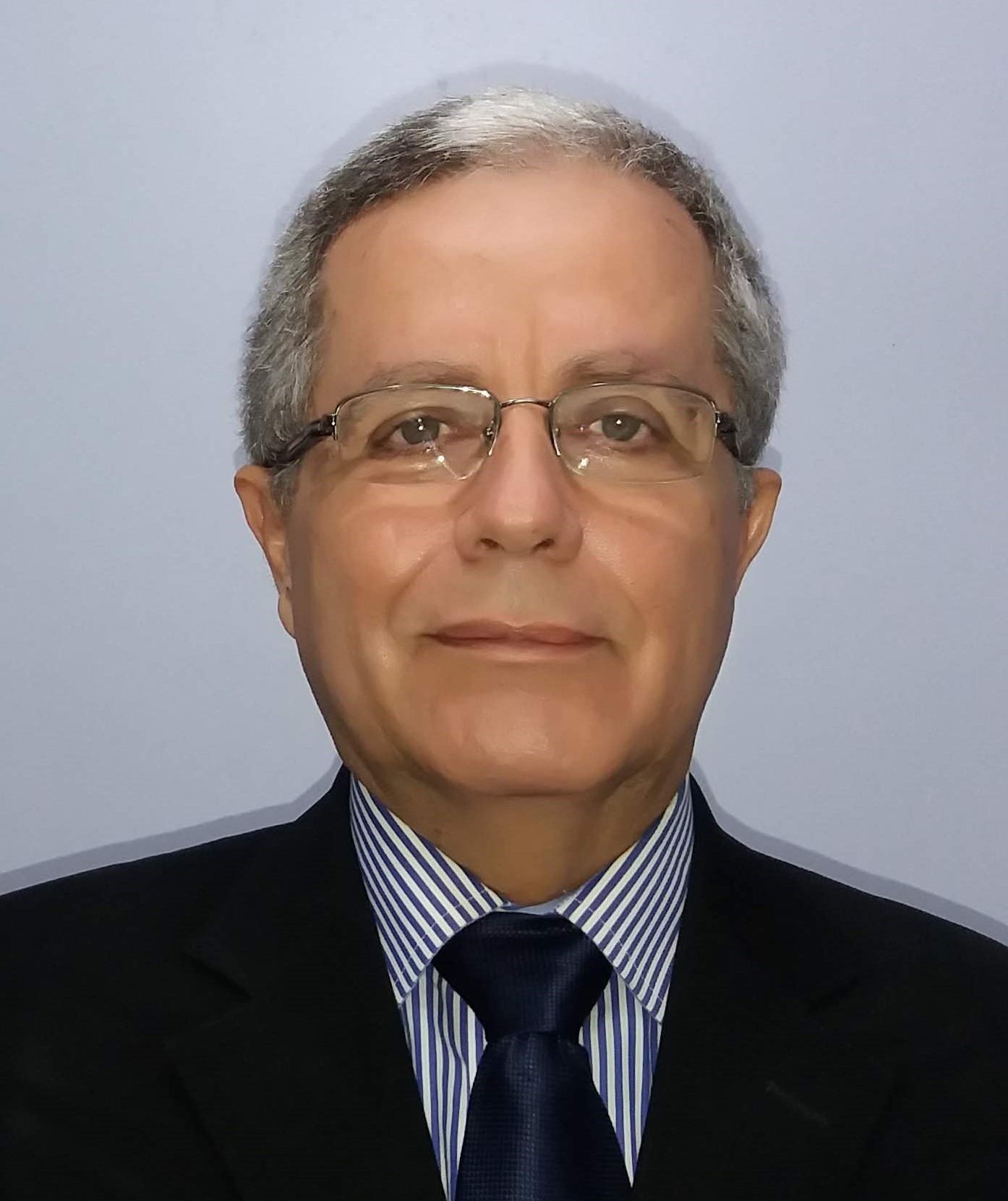}}]{Prof. Dr. Joberto S. B. Martins -}
PhD in Computer Science at Université Pierre et Marie Curie – UPMC, Paris (1986), PosDoc at ICSI/ Berkeley University (1995) and PosDoc Senior Researcher at Paris Saclay University – France (2016). International Professor at HTW (Germany) (since 2004) and Université d'Evry (France). Full Professor at Salvador University on Computer Science, Head of NUPERC and IPQoS research groups with research interests on Cognitive Management, Resource Allocation, Machine Learning, SDN/ OpenFlow, Internet of Things, Smart Grid and Smart Cities. 
\end{IEEEbiography}
\end{document}